\documentclass[twocolumn]{aastex631}

\usepackage{CJKutf8}
\usepackage{amsmath}
\usepackage{xcolor}

\newcommand{\project}[1]{\textsl{#1}}

\newcommand{\tess}{\project{TESS}}

\newcommand{\gaia}{\project{Gaia}}

\newcommand{\T}{\ensuremath{\mathrm{T}}}

\newcommand{\appropto}{\mathrel{\vcenter{
  \offinterlineskip\halign{\hfil$##$\cr
    \propto\cr\noalign{\kern2pt}\sim\cr\noalign{\kern-2pt}}}}}

\graphicspath{{./}{figures/}}

\begin{document}

\title{Six New Circumbinary Disk Occultation (CBO) Candidates from the Zwicky Transient Facility}

\author[0009-0000-6461-5256]{Zhecheng Hu (\begin{CJK*}{UTF8}{gbsn}胡哲程\end{CJK*})}
\affiliation{Department of Astronomy, Tsinghua University, Beijing 10084, China}
\correspondingauthor{Zhecheng Hu}
\email{hzc22@mails.tsinghua.edu.cn}

\author[0000-0003-4027-4711]{Wei Zhu (\begin{CJK*}{UTF8}{gbsn}祝伟\end{CJK*})}
\affiliation{Department of Astronomy, Tsinghua University, Beijing 10084, China}
\email{weizhu@mail.tsinghua.edu.cn}

\author[0009-0004-3360-3702]{Shuming Wang (\begin{CJK*}{UTF8}{gbsn}王书铭\end{CJK*})}
\affiliation{Xingming Observatory, Urumqi 830011, Xinjiang, People’s Republic of China}

\author[0000-0002-6937-9034]{Sharon Xuesong Wang (\begin{CJK*}{UTF8}{gbsn}王雪凇\end{CJK*})}
\affiliation{Department of Astronomy, Tsinghua University, Beijing 10084, China}

\begin{abstract}

Circumbinary disk occultation (CBO) systems, in which a misaligned circumbinary disk periodically obscures the central binary, provide unique probes of disk structure and dynamics. However, fewer than ten candidates with measured orbital periods were previously known. In this work, we identify six new CBO candidates, designated ZTF-CBO-1 through ZTF-CBO-6, through a systematic search of Zwicky Transient Facility (ZTF) photometry. These systems exhibit deep ($\gtrsim$1~mag) periodic dimming events with orbital periods ranging from $\sim$30 to $\sim$530~days and occultation durations spanning 32\%--77\% of their orbital periods. Such large duty cycles rule out the interpretation of circum-companion material occultation. Spectral energy distributions of ZTF-CBO-1 and ZTF-CBO-2 reveal infrared excess indicative of a dust component. TESS observations of ZTF-CBO-1 show no hours- to days-timescale variability during ingress and egress, indicating a smooth disk edge. These discoveries nearly double the known CBO sample, enabling meaningful population-level studies.

\end{abstract}


\section{Introduction}
\label{sec:intro}

Periodic disk occultation events provide a unique probe of circumstellar and circumbinary material, yet remain poorly unified due to sample scarcity. While eclipses are widely used as a standard probe to study binary stars \citep[e.g.,][]{Andersen1991_Accurate, Torres2010_Accurate} and exoplanets \citep[e.g.,][]{Charbonneau2000_Detection}, periodic occultations by dusts are far less common. These dust occultations are typically divided into two classes: CircumStellar Occultations (CSOs), exemplified by $\epsilon$~Aur \citep{Carroll1991_Interpreting}, where a circumstellar disk eclipses its host star; and CircumBinary Occultations (CBOs), exemplified by KH~15D \citep{KH15D_discover,Winn2004_KH15D,Chiang2004_KH15D}, where a circumbinary disk periodically obscures the central binary \footnote{$\epsilon$ Aur-like and KH 15D-like do not account for all dippers, especially the quasi-periodic and non-periodic ones.}. To date, only about a dozen CSO \citep{Fores-Toribio2025_ASASSN-24fw} and nine CBO candidates have known periods, preventing population-level studies of these phenomena. 

Large-scale photometric surveys have begun to expand this sample. The Optical Gravitational Lensing Experiment phase III \citep[OGLE-III;][]{Udalski2003_Optical} and the ZTF \citep[][]{Bellm2018_Zwicky, Masci2019_ZTF} have discovered disk occultation events during searches for variable stars \citep[e.g.,][]{Graczyk2011_Optical, Chen2020_Zwicky, Zhu2022_Two}, yielding at least four CSO systems (OGLE-LMC-ECL-17782, OGLE-LMC-ECL-11893, OGLE-BLG182.1.162852, and ZTF~J1852+1249; \citealt{Graczyk2011_Optical, Dong2014_OGLE-LMC-ECL-11893, Rattenbury2015_OGLE-BLG182.1.162852, Bernhard2024_ZTF}) and two CBO systems (Bernhard-1 and 2; \citealt{Zhu2022_Two,Hu2024_Bernhard2}). Additionally, the VISTA Variables in the Via Lactea Survey \citep[VVV;][]{Minniti2010_VVV} and the All-Sky Automated Survey for Supernovae \citep[ASAS-SN;][]{Shappee2014_MAN,Kochanek2017_All-Sky} have conducted targeted searches for large-amplitude dimming events \citep{Lucas2024_most,Guo2024_Spectroscopic,JoHantgen2025_ASAS-SN}, identifying one CSO candidate \citep[ASASSN-24fw;][]{Zakamska2025_ASASSN-24fw,Fores-Toribio2025_ASASSN-24fw} and one CBO candidate \citep[VVV J180502.29-242501.4;][]{Lucas2024_most}.

KH~15D remains the prototypical CBO system and demonstrates the scientific potential of such objects. In this system, a misaligned circumbinary disk periodically occults the binary stars, producing characteristic flux variations \citep{Chiang2004_KH15D,Winn2004_KH15D}. Decades of follow-up observations have yielded rich insights. Light curve modeling constrains the disk-binary mutual inclination of $\sim$15$^\circ$ and disk warping of $\sim$10$^\circ$ \citep{Winn2004_KH15D,Poon2021_Constraining}. Multi-band photometry reveals clumpy structure at the disk edge composed of large dust grains \citep{Arulanantham2016_SEEING,GarciaSotok2020_Clumps}. Spectroscopic observations has made use of the occultation process to isolate the origin of different spectral features. These observations identified water ice and methane in the disk \citep{Arulanantham2017_Untangling}, the potential [\ion{O}{1}] emission from the gaseous disk surface \citep{Fang2019_Double-peaked}, bipolar jets \citep{Deming2004_Spectroscopy, Mundt2010_Bipolar}, and pulsed accretion near pericenter passage \citep{Hamilton2012_Complex}.

The recent discovery of two additional CBO systems, Bernhard-1 and Bernhard-2 \citep{Zhu2022_Two}, demonstrates that KH~15D-like systems are not unique. Furthermore, they are later confirmed through joint radial velocity and photometric analysis \citep[][Hu et al. in prep]{Hu2024_Bernhard2}. This motivates systematic searches for additional candidates to enable population-level studies.

In this paper, we report six new CBO candidates identified through a systematic search of ZTF photometric data. These candidates exhibit periodic deep ($\gtrsim$ one magitude), long-duration ($\gtrsim$ 30\% of the orbital periods) dimming events consistent with circumbinary disk occultation. Three candidates (ZTF-CBO-1, 3, and 4) were initially identified by co-author S.M. Wang through an independent search as an amateur astronomer, while subsequent vetting and complementary searches yielded the remaining three.

This paper is organized as follows. Section~\ref{sec:candidate-search} describes the search process and cross-matching with other observations. Section~\ref{sec:model} presents the light curve and spectral energy distribution models. Section~\ref{sec:systems} reports the detailed analysis of each system. Section~\ref{sec:do-summary} provides an overview of known periodic deep-and-long eclipsing events, and Section~\ref{sec:conclusion} discusses our conclusions.

\begin{deluxetable*}{l cc cc cc cc cc cc}
\rotate
\tablewidth{0pt}
\tabletypesize{\small}
\tablecaption{Basic and SED information of the six circumbinary disk occultation candidates.}
\label{tab:info}
\label{tab:mag}
\tablehead{
Parameter / Filter & \multicolumn{2}{c}{ZTF-CBO-1} & \multicolumn{2}{c}{ZTF-CBO-2} & \multicolumn{2}{c}{ZTF-CBO-3} & \multicolumn{2}{c}{ZTF-CBO-4} & \multicolumn{2}{c}{ZTF-CBO-5} & \multicolumn{2}{c}{ZTF-CBO-6}
}
\startdata
\gaia\ DR3 ID & \multicolumn{2}{c}{3439543092062056576} & \multicolumn{2}{c}{4094461626879471616} & \multicolumn{2}{c}{4145750923747039104} & \multicolumn{2}{c}{2058213238096217600} & \multicolumn{2}{c}{4104926233160064128} & \multicolumn{2}{c}{3426641014604832128} \\
RA$_{\mathrm{J2000}}$ & \multicolumn{2}{c}{$06^{\rm h}28^{\rm m}53\fs10$} & \multicolumn{2}{c}{$18^{\rm h}16^{\rm m}27\fs04$} & \multicolumn{2}{c}{$18^{\rm h}16^{\rm m}39\fs94$} & \multicolumn{2}{c}{$20^{\rm h}05^{\rm m}10\fs39$} & \multicolumn{2}{c}{$18^{\rm h}35^{\rm m}26\fs56$} & \multicolumn{2}{c}{$06^{\rm h}04^{\rm m}07\fs56$} \\
Dec$_{\mathrm{J2000}}$ & \multicolumn{2}{c}{$+33\arcdeg33\arcmin56\farcs6$} & \multicolumn{2}{c}{$-19\arcdeg47\arcmin21\farcs1$} & \multicolumn{2}{c}{$-15\arcdeg44\arcmin33\farcs9$} & \multicolumn{2}{c}{$+33\arcdeg42\arcmin31\farcs6$} & \multicolumn{2}{c}{$-12\arcdeg51\arcmin40\farcs3$} & \multicolumn{2}{c}{$+25\arcdeg35\arcmin00\farcs0$} \\
Parallax (mas) & \multicolumn{2}{c}{$0.67 \pm 0.04$} & \multicolumn{2}{c}{$0.56 \pm 0.21$} & \multicolumn{2}{c}{$0.64 \pm 0.11$} & \multicolumn{2}{c}{$0.4 \pm 0.6$} & \multicolumn{2}{c}{$0.41 \pm 0.16$} & \multicolumn{2}{c}{$0.0 \pm 0.6$} \\
$P$ (days) & \multicolumn{2}{c}{$529.32 \pm 0.03$} & \multicolumn{2}{c}{$74.454 \pm 0.005$} & \multicolumn{2}{c}{$151.380 \pm 0.005$} & \multicolumn{2}{c}{$30.2566 \pm 0.0004$} & \multicolumn{2}{c}{$61.2764 \pm 0.0022$} & \multicolumn{2}{c}{$149.615 \pm 0.012$} \\
$t_{\mathrm{in}}$ (MJD) & \multicolumn{2}{c}{$58725.27 \pm 0.07$} & \multicolumn{2}{c}{$59009.69 \pm 0.09$} & \multicolumn{2}{c}{$58638.94 \pm 0.05$} & \multicolumn{2}{c}{$58334.221 \pm 0.017$} & \multicolumn{2}{c}{$59030.304 \pm 0.022$} & \multicolumn{2}{c}{$59177.55 \pm 0.06$} \\
$t_{\mathrm{out}}$ (MJD) & \multicolumn{2}{c}{$58895.69 \pm 0.04$} & \multicolumn{2}{c}{$59055.09 \pm 0.05$} & \multicolumn{2}{c}{$58693.254 \pm 0.028$} & \multicolumn{2}{c}{$58351.652 \pm 0.015$} & \multicolumn{2}{c}{$59055.22 \pm 0.05$} & \multicolumn{2}{c}{$59295.14 \pm 0.12$} \\
$v_{\mathrm{in}}$ ($R_{\star}$/day) & \multicolumn{2}{c}{$0.0619 \pm 0.0005$} & \multicolumn{2}{c}{$0.193 \pm 0.006$} & \multicolumn{2}{c}{$0.0953 \pm 0.0009$} & \multicolumn{2}{c}{$0.709 \pm 0.018$} & \multicolumn{2}{c}{$2.91 \pm 0.11$$^\dagger$} & \multicolumn{2}{c}{$0.236 \pm 0.008$} \\
$v_{\mathrm{out}}$ ($R_{\star}$/day) & \multicolumn{2}{c}{$0.0852 \pm 0.0009$} & \multicolumn{2}{c}{$0.205 \pm 0.004$} & \multicolumn{2}{c}{$0.1126 \pm 0.0007$} & \multicolumn{2}{c}{$1.46 \pm 0.05$} & \multicolumn{2}{c}{$0.628 \pm 0.024$} & \multicolumn{2}{c}{$0.240 \pm 0.012$} \\
\hline
\multicolumn{13}{c}{\textit{SED magnitudes}} \\ \\
& in & out & in & out & in & out & in & out & in & out & in & out \\
$g$ & 16.692(5) & 15.5156(28) & \dots & 19.89(5) & 19.72(4) & 17.525(8) & \dots & \dots & 19.96(4) & 19.006(19) & 21.97(18) & 20.32(6) \\
$r$ & 16.121(4) & 15.0625(20) & 20.49(7) & 18.199(17) & \dots & 16.679(5) & \dots & \dots & 18.697(17) & 18.080(12) & 20.64(6) & \dots \\
$i$ & \dots & 14.8678(19) & 19.56(3) & 17.085(7) & 17.606(8) & 16.120(3) & 20.53(12) & 19.084(25) & 18.107(12) & 17.249(7) & 19.68(4) & 18.569(15) \\
$z$ & 16.00(3) & 14.7653(29) & 19.05(4) & 16.392(8) & 17.205(11) & 15.837(7) & 19.78(6) & 18.036(21) & \dots & 17.298(11) & 19.33(5) & 18.056(16) \\
$y$ & 15.88(3) & 14.695(5) & 18.87(11) & 15.957(13) & 16.851(12) & 15.622(11) & 19.42(14) & 17.78(6) & \dots & 16.453(21) & 18.95(8) & \dots \\
$J$ & 14.96(4) & \dots & \dots & 14.312(19) & \dots & 14.39(13) & \dots & \dots & \dots & \dots & \dots & \dots \\
$H$ & 14.51(5) & \dots & \dots & 13.232(20) & \dots & \dots & \dots & \dots & \dots & \dots & \dots & \dots \\
$K_{s}$ & 14.43(7) & \dots & \dots & 12.825(19) & \dots & \dots & \dots & \dots & \dots & \dots & \dots & \dots \\
$W_{1}$ & \dots & 13.19(5) & 15.0(5) & 12.85(10) & \dots & \dots & \dots & \dots & \dots & \dots & 16.0(3) & 15.59(18) \\
$W_{2}$ & \dots & 13.06(8) & \dots & 13.20(22) & \dots & \dots & \dots & \dots & \dots & \dots & 15.2(4) & 15.2(4) \\
$W_{3}$ & \dots & 11.0(4) & \dots & 7.95(17) & \dots & \dots & \dots & \dots & \dots & \dots & \dots & \dots \\
$W_{4}$ & \dots & 7.07(28) & \dots & 5.27(13) & \dots & \dots & \dots & \dots & \dots & \dots & \dots & \dots \\
\enddata
\tablecomments{
$^{\dagger}$: The inferred $v_{\mathrm{in}}$ for ZTF-CBO-5 may be affected by systematics and lies near the upper bound of the adopted sampling range (3 $R_{\star}\,\mathrm{day}^{-1}$).
}
\end{deluxetable*}

\section{Candidate Search and Data Collection}
\label{sec:candidate-search}

We applied a specialized search for CBO signatures. When the primary star is periodcially occulted, CBO systems typically exhibit deep ($\gtrsim 1$ mag) and long ($\gtrsim 30\%$ of the orbital period) periodic dimmings, while the photometric baseline typically remains stable. Though long-term secular variability is also observed in KH 15D and Bernhard-2 \citep{Winn2004_KH15D, Zhu2022_Two, Hu2024_Bernhard2}, we do not expect this to affect our search considering the sparse sampling of ZTF observations. Our multi-stage search process combined catalog pre-selection, periodic occultation detection, and visual vetting, inspired by the strategy of S.M. Wang and \citet{Zhu2022_Two}.

We begin by pre-selecting variable stars from the Automatic Learning for the Rapid Classification of Events \citep[ALeRCE;][]{Forster2021_Automatic} alert broker. Using the ALeRCE Light Curve classifier \citep{Sanchez-Saez2021_Alert}, we queried sources classified as periodic variables or young stellar objects (YSOs) with the highest probability. CBO systems are periodic and typically occur in YSOs, making these catalogs a natural starting point. This selection includes over 1.5 million periodic variables and 0.5 million YSOs. This sample is approximately twice the size of that analyzed by \citet{Zhu2022_Two}, enabling the discovery of additional CBO candidates.

For each source, we retrieved ZTF DR23 $g$-, $r$-, and $i$-band light curves within 1.5 arcsec using SNAD \citep{Malanchev2023_SNAD}. We retained only observations with good quality flag, i.e., \texttt{catflags} $< 32678$ \citep{Masci2019_ZTF}, and applied three cuts: (1) at least 100 total observations; (2) a baseline exceeding 100 days; and (3) a peak-to-dip amplitude greater than 1 mag in at least one band. Fewer than 0.5 million targets passed these cuts.

We then searched for long-duration periodic occultations using the box least-squares (BLS) periodogram in \texttt{astropy} \citep{astropy:2013,astropy:2018,astropy:2022}. We scanned periods from 10 to 1000 days in 0.5-day steps, with occultation durations spanning 5\% to 50\% of the period. We considered a periodic signal significant if it satisfied the following empirical criteria: (1) maximum BLS power $> 10^{4}$; (2) a maximum-to-median power ratio $> 2$; (3) the best period $> 15$ days to fliter eclipsing binaries; and (4) the best period outside the one-year alias window ($|P - 365.25| > 10$ days).

To distinguish occultations from smooth variability, we compared a periodic box model to a third-order polynomial model. Both models were fitted by minimizing $\chi^{2}$ using the Nelder--Mead algorithm in \texttt{scipy} \citep{2020SciPy-NMeth}. We renormalized uncertainties to yield $\chi^{2}_{\rm box}/{\rm dof} = 1$ for the box model and required the polynomial model to have reduced $\chi^{2} > 11$. This procedure yielded 470 candidates.

We refined periods using phase-dispersion minimization (PDM) over 10--1000 days with 0.1-day steps, then phase-folded and visually inspected each candidate. We verified stable photometry both outside and during occultation, and checked for consistent occultation depth and shape across cycles to exclude long-period variables (LPV).

This search identified eight CBO candidates: the previously reported Bernhard-1 and Bernhard-2, plus six new systems designated ZTF-CBO-1 through ZTF-CBO-6. KH 15D was not recovered because ALeRCE classifies it as neither a periodic variable nor a YSO, and therefore its ZTF data were not retrieved. We cross-matched the six new candidates with Pan-STARRS \citep[PS $g$, $r$, $i$, $z$, $y$;][]{Chambers2016_PS1, Flewelling2020_Pan-STARRS1}, 2MASS \citep[$J$, $H$, $K_s$;][]{Skrutskie2006_2MASS}, and WISE \citep[$W$1--4;][]{Wright2010_WISE} within $1.5''$ to characterize their spectral energy distributions (SED). Additionally, we searched \gaia DR3 \citep{GaiaCollaboration2016_Gaia, Gaia2023_DR3} within $3''$ to verify that the brightness variations originate from the intended targets rather than nearby contaminating sources. Detailed analysis of these properties are discussed case by case in Section \ref{sec:systems}.

ZTF-CBO-1, the brightest candidate, falls within the \tess \ \citep{Ricker2015_TESS} detectable range. \tess \ observed it in sectors 20, 43--45, 60, and 71--73. Most of these sectors sample only the outside or during occultation phases, but sector 60 captured part of the egress and sector 73 captured part of the ingress. We extracted light curves from the full-frame images using the TESS-Gaia Light Curve pipeline \citep[\texttt{tglc};][]{Han2023_TESS-Gaia}. A $90 \times 90$ pixel cutout is applied to estimate the empirical point-spread function and correct for background contamination. We did not apply detrending, as recommended by \citet{Han2023_TESS-Gaia} for long-term variables.

\section{Modeling}
\label{sec:model}

\begin{figure}
\includegraphics[width=1.0\linewidth]{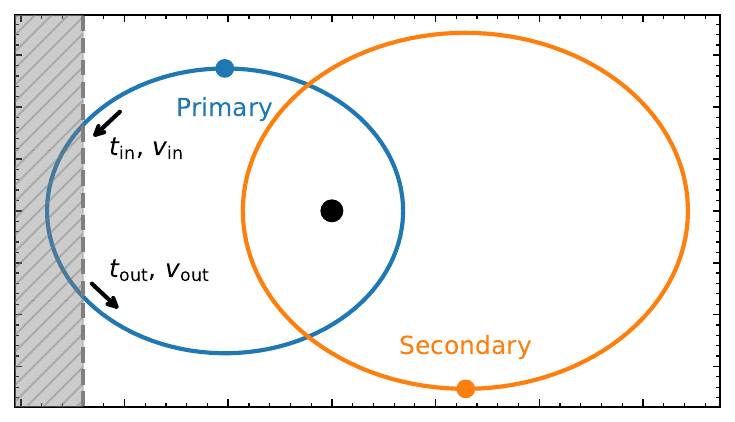}
\caption{Schematic diagram of an eccentric binary system with a misaligned circumbinary disk. The shaded region represents the disk, which obscures part of the orbit of the primary star. When the primary star passes behind the disk once per orbital period, the system brightness drops significantly.}
\label{fig:schematic}
\end{figure}

We model ZTF-CBO-1 to 6 as binaries periodically occulted by their circumbinary disks. We assume the secondary star remains visible throughout the orbit. A schematic of this model is shown in Figure~\ref
{fig:schematic}.

\begin{figure*}
\centering
\includegraphics[width=0.9\linewidth]{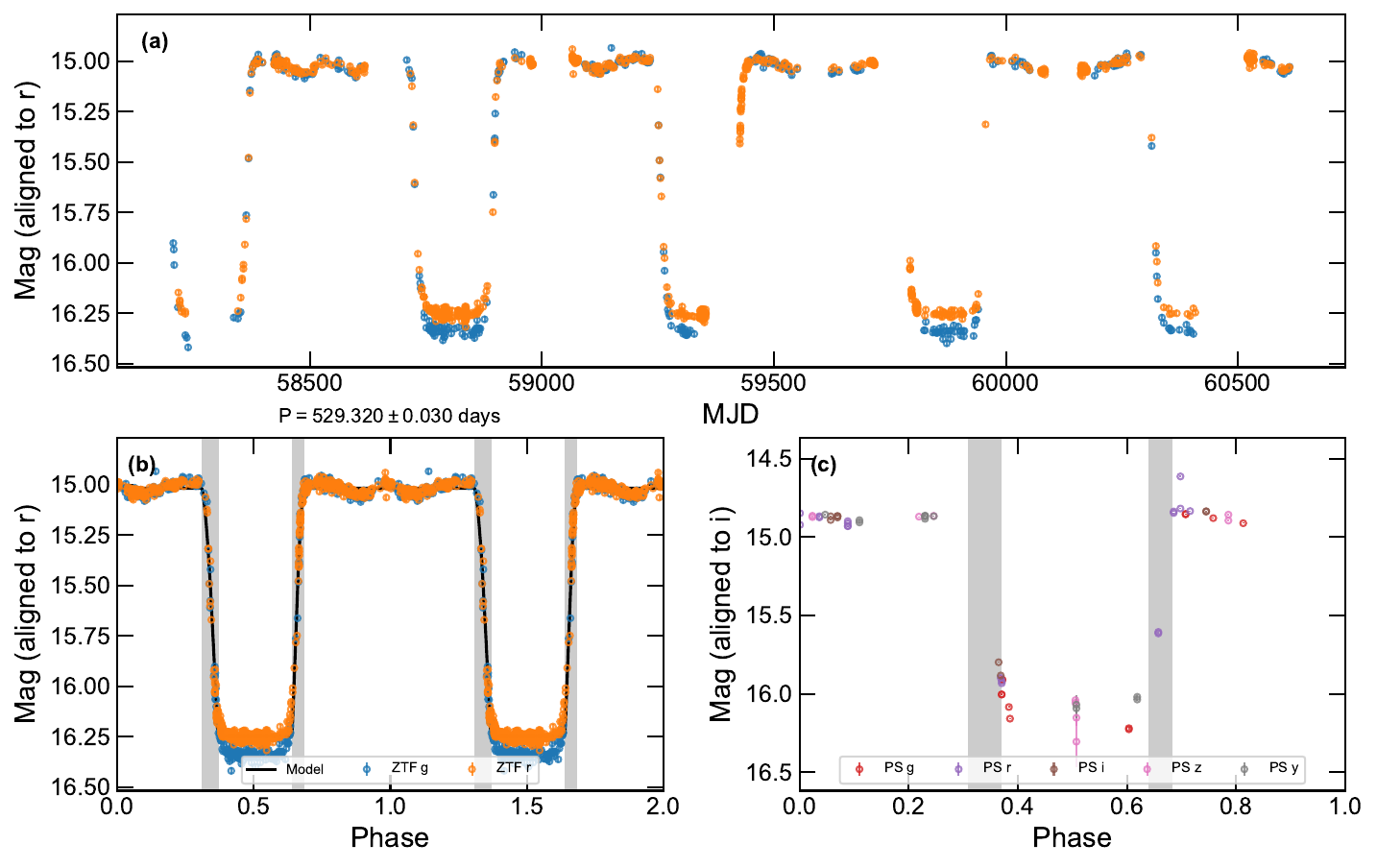}
\caption{Three views of the ZTF-CBO-1 light curve. Panel (a): original ZTF light curve spanning multiple years. Panel (b): phase-folded light curve binned to 0.1 day per point, repeated for two periods to show both the full baseline and the complete occultation. Panel (c): phase-folded PS light curve. Light curves in each band are shifted vertically to align with the $r$ and $i$ bands for ZTF and PS, respectively. Grey regions in panels (b) and (c) mark the model predicted ingress and egress phases.}
\label{fig:CBO1_P530}
\end{figure*}

\subsection{Light Curve Model}

We adopt the opaque screen model from \citet{Zhu2022_Two} to characterize the occultation events. This model assumes a sharp, opaque disk edge and no limb darkening accross the stellar photospehere. The occultation is then fully characterized by the motion of the star relative to the disk edge.

The visible fraction $\eta(t)$ of the primary star depends on five geometric parameters: the orbital period $P$, the ingress and egress mid-times, i.e., $t_{\rm in}$ and $t_{\rm out}$, and the projected crossing velocities for ingress ($v_{\rm in}$) and egress ($v_{\rm out}$). The $\eta(t)$ is given by
\begin{equation}
\eta(t) = \begin{cases} 
1 & , x \le -1 \\
\frac{1}{2} - \frac{1}{\pi} \left[ x\sqrt{1-x^2} + \arcsin x \right] & , -1 < x < 1 \\
0 & , x \ge 1
\end{cases}
\end{equation}
where the position $x$ encodes the location of the star relative to the disk edge in unit of the stellar radius:
\begin{equation}
x = \begin{cases} 
v_{\text{in}}(t - t_{\text{in}}) & , \text{ingress} \\
v_{\text{out}}(t_{\text{out}} - t) & , \text{egress}
\end{cases}
.
\end{equation}
Let $F_1$ and $F_2$ denote the flux contributions from the primary and secondary stars, respectively. Since the secondary remains visible, the total observed flux is
\begin{equation}
F(t) = F_1 \cdot \eta(t) + F_2 .
\label{eq:f1f2}
\end{equation}
The primary star visible fraction $\eta$ is a dimensionless quantity determined solely by the occultation process. By normalizing the measured flux with $F_1$ and $F_2$, we isolate $\eta$ from the individual stellar contributions. This normalization enables a direct comparison of occultation profiles observed with different instruments and filters.

We fit the ZTF photometry using a two-stage approach. For each trial set of geometric parameters $(P, t_{\rm in}, t_{\rm out}, v_{\rm in}, v_{\rm out})$, we solve for $F_1$ and $F_2$ via linear regression and evaluate the likelihood. We then sample the posterior with the \texttt{emcee} Markov chain Monte Carlo (MCMC) sampler \citep{Foreman-Mackey2013_emcee}. The best-fit parameters are reported in Table~\ref{tab:info}.

\begin{figure*}
\centering
\includegraphics[width=0.8\linewidth]{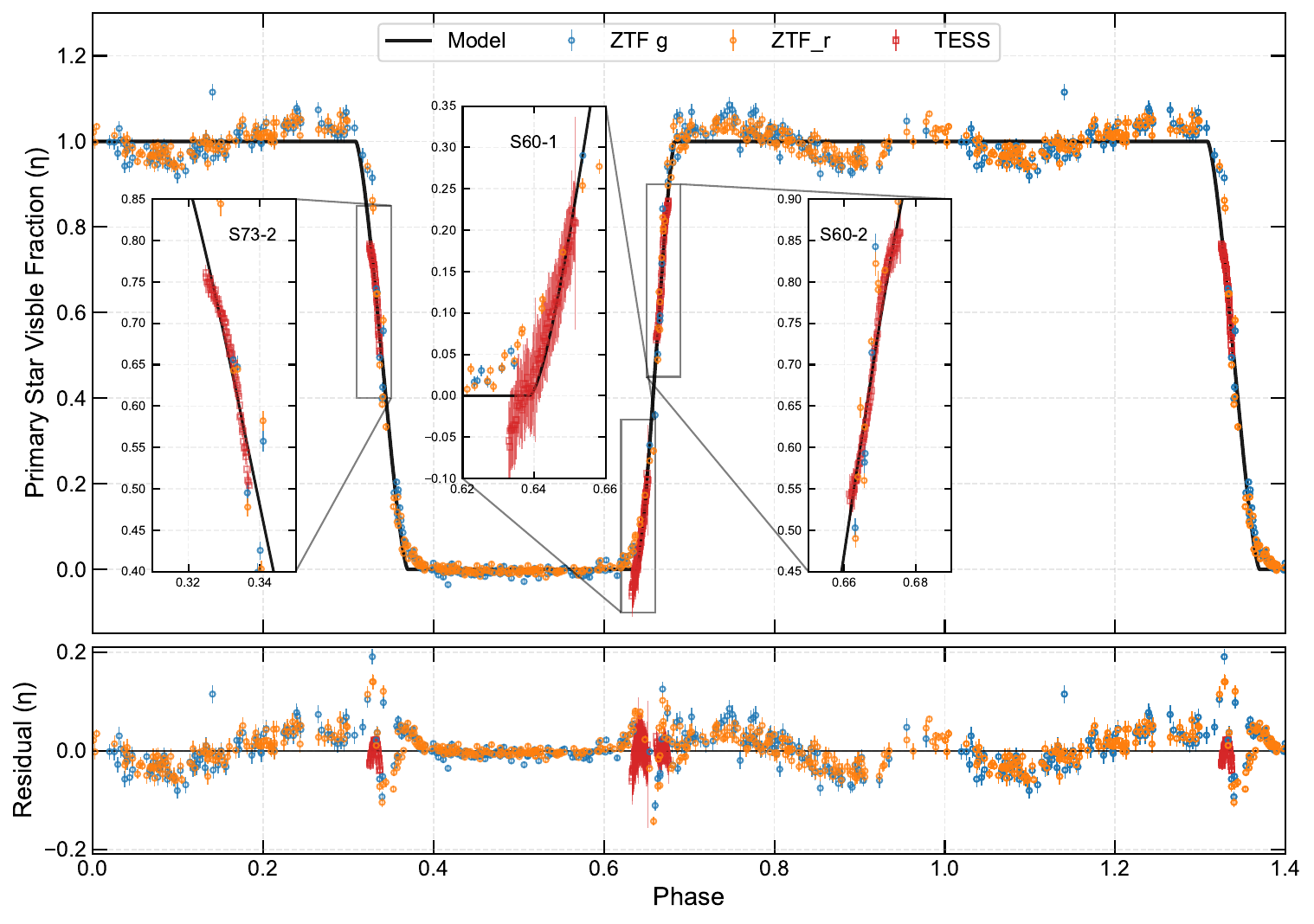}
\caption{Visible fraction $\eta$ of the primary star in ZTF-CBO-1, combining ZTF and \tess \ observations. The light curve from phase 1.0 to 1.4 repeats the segment from phase 0.0 to 0.4 to show both the full outside and during occultation baseline. Inset panels highlight the light curve covered by \tess, which are labeled by sector and dataset number. No significant fluctuations are detected on timescales from hours to days. After egress, the source first reaches a maximum, followed with a fading by $\approx 0.07$~mag, then returns to a second maximum near phase 1.0. This baseline modulation appears symmetric about phase 1.0.}
\label{fig:CBO1_P530_zoom}
\end{figure*}

\subsection{TESS Light Curve Model}
\label{subsec:tess_model}

The \tess \ data for ZTF-CBO-1 requires a special treatment. Due to the low resolution of \tess \ and the faintness of the source, the flux calibration varies between sectors. Since \tess \ communicates with Earth once per sector and the calibration may change afterward, we divide the data into datasets based on observation continuity. The calibration variations prevent us from 
directly fitting the model on the combined \tess \ data. Our primary goal is thus to detect short-term variability with \tess.

The observed \tess \ flux variations differ from ZTF predictions due to contamination. During egress in sector 60, the \tess \ flux increased by $\sim$75\%, which is substantially smaller than the $\sim$190\% predicted from the ZTF $r$-band model. During ingress in sector 73, the flux decreased by $\sim$30\%, consistent with the ZTF prediction. Despite the contamination, these variations confirm that ZTF-CBO-1 contributes significantly to the total PSF flux.

We restrict our analysis to datasets that overlap with the ingress or egress phases. Both datasets in sector 60 cover part of the egress and the second dataset in sector 73 captures part of the ingress. For each dataset, we fix the geometric parameters at their ZTF best-fit values and fit only $F_1$ and $F_2$ as defined in Equation~\ref{eq:f1f2}. Here $F_1$ and $F_2$ account for contamination rather than intrinsic stellar flux. We use these parameters to convert \tess \ flux into the primary star visible fraction $\eta(t)$. A detailed analysis of the \tess \ data is shown in Section~\ref{subsec:CBO1}.

\subsection{SED Model}

To measure the individual stellar fluxes, we extract phase-resolved SEDs from PS, 2MASS, and WISE photometry. Observations are grouped into three phases: during occultation, outside occultation, and ingress/egress. The median magnitudes from the outside and during occultations represent the total binary flux and secondary flux, respectively. We include a maximum photometric accuracy of 0.03 mag to account for potential systematics between different surveys, which is added in quadrature to the reported photometric uncertainty. The SED measurements are listed in Table~\ref{tab:mag}.

We model the SEDs based on the \texttt{isochrones} package \citep{Morton2015_isochrones}, which interpolates on MIST \citep{Choi2016_MIST,Dotter2016_MIST} to determine stellar parameters. The magnitude in a certain band $B$ for a single star is given by
\begin{equation}
    m_{\rm B} = f( {\rm EEP}, \log {\rm age}, [{\rm Fe/H}], d, A_V)
\end{equation}
where EEP is the equivalent evolutionary phase \citep{Dotter2016_MIST}, $d$ is the distance, and $A_V$ is the extinction. We adopt the standard \citet{Cardelli1989_extinction} extinction law with $R_V = 3.1$. The two stars share all parameters except EEP to allow for different masses and evolutionary states. We simultaneously model the SEDs measured outside and during occultation as the total and secondary fluxes.

We apply this analysis to ZTF-CBO-1 and ZTF-CBO-2, which have both the 2MASS and WISE data available. We use \gaia \ parallax shown in Table~\ref{tab:info} as a distance prior. The W3 and W4 bands are excluded as they are affected by infrared excess. The best-fit model is obtained by optimizing the total log-likelihood of all measurements, the \texttt{emcee} sampler \citep{Foreman-Mackey2013_emcee} is used to sample the posterior. However, the SED modeling for both sources suffers from multimodal degeneracy, so we do not report detailed stellar properties.

\section{DISK-OCCULTATION SYSTEMS}
\label{sec:systems}

\begin{figure*}
\centering
\includegraphics[width=1.0\linewidth]{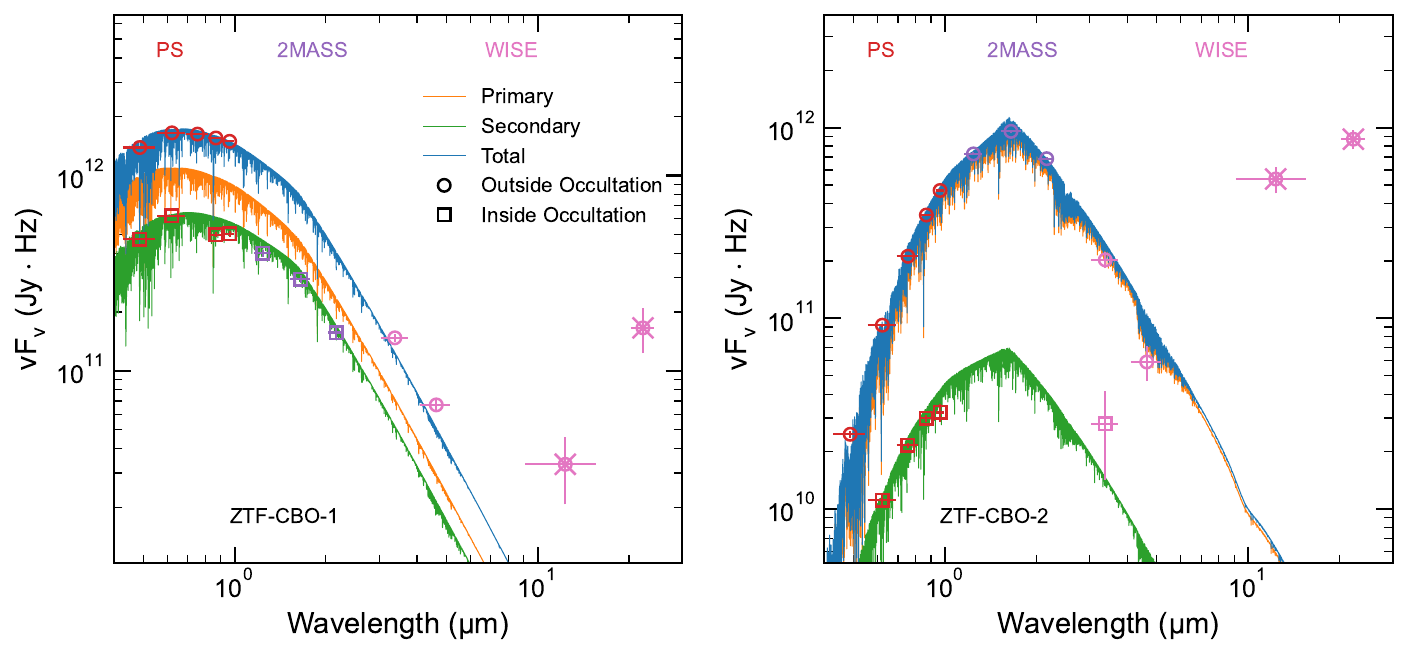}
\caption{SEDs and best-fit stellar models for ZTF-CBO-1 (left) and ZTF-CBO-2 (right). Photometric measurements from PS, 2MASS, and WISE are shown, with circles and squares indicating observations obtained outside of and during occultation, respectively. Orange and green curves show the primary and secondary stellar components, respectively, and the blue curve shows the combined model. ZTF-CBO-1 and ZTF-CBO-2 are the only targets in our sample with an infrared excess detected in WISE W3 and W4 bands. Observations in these two bands are not used in the fitting process, as indicated by the cross on the marker.}
\label{fig:SED}
\end{figure*}

\subsection{ZTF-CBO-1}
\label{subsec:CBO1}

ZTF-CBO-1, the brightest candidate and the only one with \tess \ coverage, is also the best-characterized object in our sample. The system has an orbital period of $\sim$530 days and an occultation duration of $\sim$170 days ($\sim$32\% of the period). Figure~\ref{fig:CBO1_P530} shows the ZTF and PS light curves. To examine the occultation morphology in detail using multiple datasets, we convert both ZTF and \tess \ light curves to the primary star visible fraction $\eta$, as detailed in Section~\ref{subsec:tess_model}. Figure~\ref{fig:CBO1_P530_zoom} displays the phase-folded $\eta$ curves and their residuals.

The out-of-occultation baseline has a repeatable, roughly symmetric modulation around phase 1.0, as shown in Figure~\ref{fig:CBO1_P530_zoom}. After egress, the system brightens to a maximum near phase 0.75, followed with a fading by $\sim$0.07 mag toward phase 0.9, and then rises to a second maximum near phase 1.0. This pattern is mirrored after phase 1.0, with a dip near phase 1.1 followed by rebrightening to another maximum near phase 1.25.

A single-edge opaque-screen model cannot fully explain this behavior. As the phase approaches 1, the primary star is near pericenter and farthest from the occulting edge, as shown in Figure \ref{fig:schematic}. Simultaneously, the secondary star is closest to the edge and may be partially obscured, creating a dip in the out-of-occultation baseline as seen in KH 15D \citep[e.g.,][]{Winn2004_KH15D}. However, the subsequent return to maximum brightness near phase 1.0 is difficult to reproduce with a single edge. A more complex structure, such as a second dust lane, might explain the modulation, but more observations are needed to further motivate this.

The \tess \ data show no short-timescale variability down to $\sim 10 \%$ from hours to several days during ingress and egress, as shown in Figure~\ref{fig:CBO1_P530_zoom}. This supports an smooth occulting edge without small dust clumps. Assuming an eccentricity of $\sim$0.5, and a euqal solar-mass binary as suggested by the SED fitting, the transverse velocity near apocenter is approximately one solar radius per day. The absence of rapid variability in $\eta(t)$ implies that any substructures near the disk edge are significantly larger than the stellar radii.

Phase-resolved SED fitting reveals two possible evolutionary scenarios. Using PS, 2MASS, and WISE photometry, we modeled the SEDs inside and outside of occultation, as shown in the left panel of Figure~\ref{fig:SED}. The fitting identifies a young-star solution with an age of $\sim 50$ Myr, although alternative solutions cannot be firmly excluded. Future spectroscopic observations are needed to confirm its young nature.


\begin{figure*}
\centering
\includegraphics[width=0.9\linewidth]{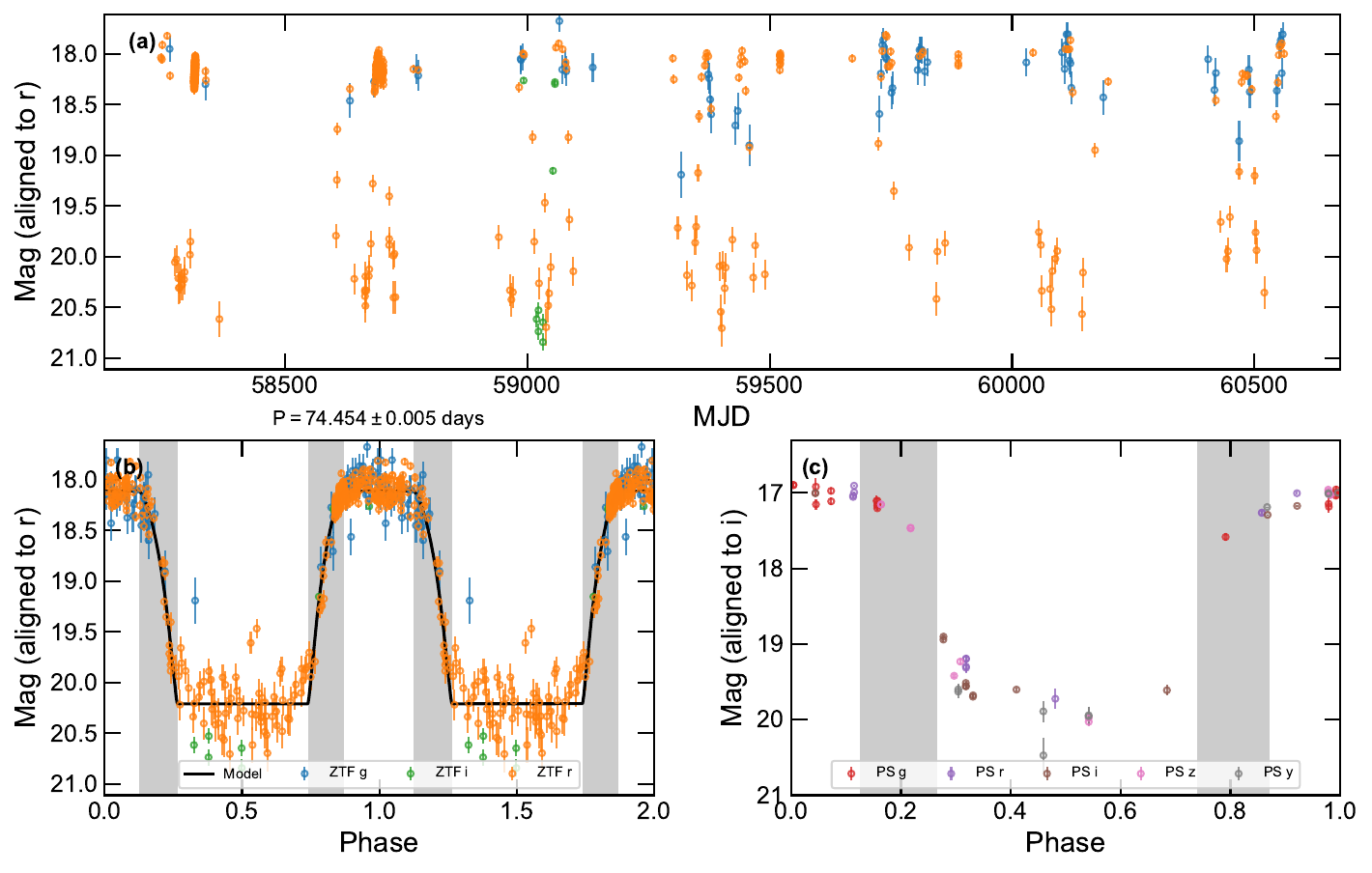}
\caption{Same as Figure \ref{fig:CBO1_P530}, but for ZTF-CBO-2.}
\label{fig:DOC2_yso_P74}
\end{figure*}

\begin{figure}[t!]
\centering
\includegraphics[width=1.0\linewidth]{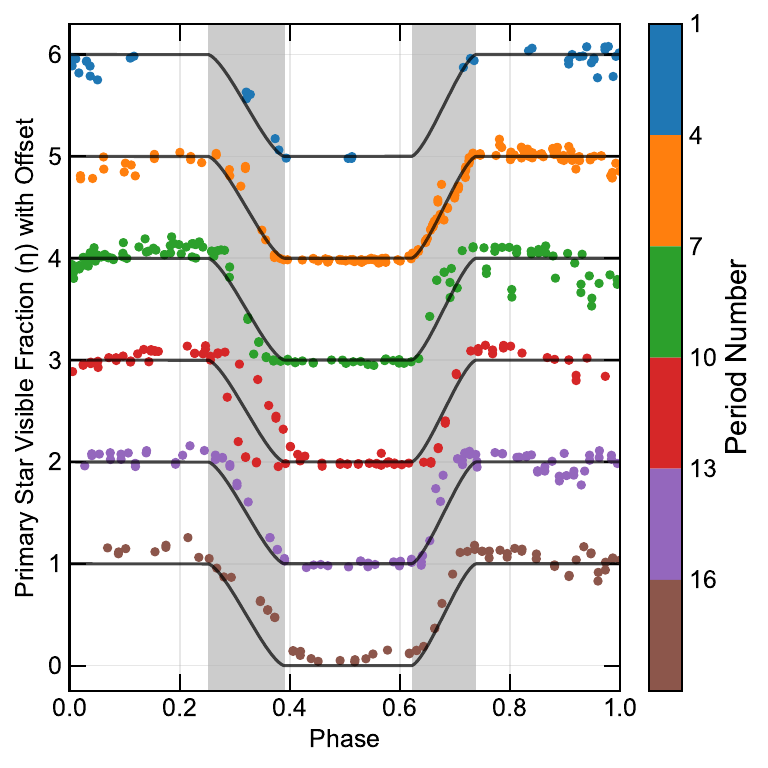}
\caption{The primary star visible fraction $\eta$ for ZTF-CBO-3, with data points color-coded and vertically shifted according to period number. The best-fit model is shown as black lines, and the predicted ingress and egress phases are marked by grey bands. Observations in the first six periods generally follow the model, while observations from periods 7 to 16 exhibit greater scatter. This scatter is especially significant during ingress and egress. The occultation depth in period 16 is likely shallower than that in the other periods.}
\label{fig:lc_DOC3_zoom}
\end{figure}

\subsection{ZTF-CBO-2}

ZTF-CBO-2 exhibits the strongest photometric variability in our sample, with a peak-to-dip amplitude exceeding 2.5 mag in both ZTF and Pan-STARRS $i$ bands. The system has an orbital period of $\sim$74 days and an occultation duration of $\sim$46 days ($\sim$63\% of the cycle), as shown in Figure \ref{fig:DOC2_yso_P74}.

SED modeling suggests a distant, evolved binary with a dusty disk. A two-component stellar model favors an evolved binary at $\gtrsim$10 kpc, composed of a red giant and a subgiant. No young-star solution with age $\lesssim$100 Myr provides a comparably good fit. The broadband SED reveals an infrared excess indicative of circumstellar dust, as shown in Figure~\ref{fig:SED}.

We also tested a single-star model, motivated by the deep occultation suggesting that the primary dominates the out-of-occultation flux. This yields a solution of a very young ($\sim 0.1$ Myr), low-mass ($\sim 0.3$) star with high extinction ($A_V \approx 3.6$). However, models at such early evolutionary phases are generally less reliable.

Despite the SED modeling ambiguities, \gaia \ and 2MASS provide two key constraints. First, the \gaia \ parallax places the source beyond 800 pc with 3-sigma significance. Second, the out-of-occultation 2MASS photometry shows the SED peaking near 1.5 $\mu$m, implying an effective temperature of $\sim$3000 K assuming negligible reddening. Together, these constraints favor a distant giant. Future spectroscopy to measure $\log g$ will be essential to determine the true nature of ZTF-CBO-2.

\subsection{ZTF-CBO-3}

ZTF-CBO-3 has significant variability in its occultation profile. The system has an orbital period of $\sim$151 days and an occultation duration of $\sim$54 days (36\% of the cycle). The phase-folded light curves preserve the overall occultation shape, but the ingress and egress phases exhibit considerably larger scatter than the other five candidates. A dip of varying depth also appears on the out-of-occultation baseline near phase 1. The ZTF and PS light curves presented in Appendix \ref{app:A} display these variations, which suggest that the occulting structure evolves over time.

The ingress and egress morphology evolves significantly across observing seasons. Figure~\ref{fig:lc_DOC3_zoom} presents the primary star visible fraction $\eta$, color-coded and shifted by period number to illustrate these variations. Observations from the first six periods (blue and orange points) are stable and generally follow the best-fit model indicated by the black line. In contrast, observations from periods 7--16 show significant deviations. The ingress timing changes rapidly between periods 10 and 12 (red points), and the occultation in period 16 (brown points) appears shallower than expected.

These variations likely result from a dynamic disk edge involving both disk precession and structural inhomogeneities. The cycle-to-cycle changes resemble the precession-induced evolution observed in KH 15D and Bernhard-2 \citep{Winn2004_KH15D, Zhu2022_Two, Hu2024_Bernhard2}. However, the precession alone cannot explain the rapid variations from periods 10 to 12. These fast changes may instead be linked to non-uniformities, such as dust clumps or spirals, at the inner edge of the circumbinary disk. Such structures would orbit at the local Keplerian velocity, potentially causing the observed short-term variations in ingress timing and depth. 

ZTF-CBO-3, ZTF-CBO-4, and ZTF-CBO-6 exhibit a symmetric dip near the center of the out-of-occultation baseline at phase 1, where the $r$-band flux drops by 5\%--7\%. These dips likely arise from the orbital motion of the secondary star. At phase 1.0, the binary components are near pericenter, bringing the secondary closest to the occulting disk edge as shown in Figure \ref{fig:schematic}. This proximity may cause a partial occultation, which is observed in KH 15D \citep{Winn2004_KH15D}. Future photometric monitoring is necessary to fully characterize the dip and the variation of the main occultation profile.

\subsection{ZTF-CBO-4}

ZTF-CBO-4 is the shortest-period system in our sample, with an orbital period of $\sim$30 days and an occultation duration of $\sim$17 days (57\% of the period). The ZTF and PS light curves are shown in Appendix \ref{app:A}. The source displays deep flux variations exceeding 2 mag in both ZTF $r$ and $i$ bands. The flux variations in $g$ band remains unknown, as the source falls below the ZTF detection limit in $g$ band during the dark phase, leaving no $g$-band data available there. We also observe a symmetric dip of approximately 6\% near the center of the out-of-occultation baseline.

\begin{figure}[t!]
\centering
\includegraphics[width=1.0\linewidth]{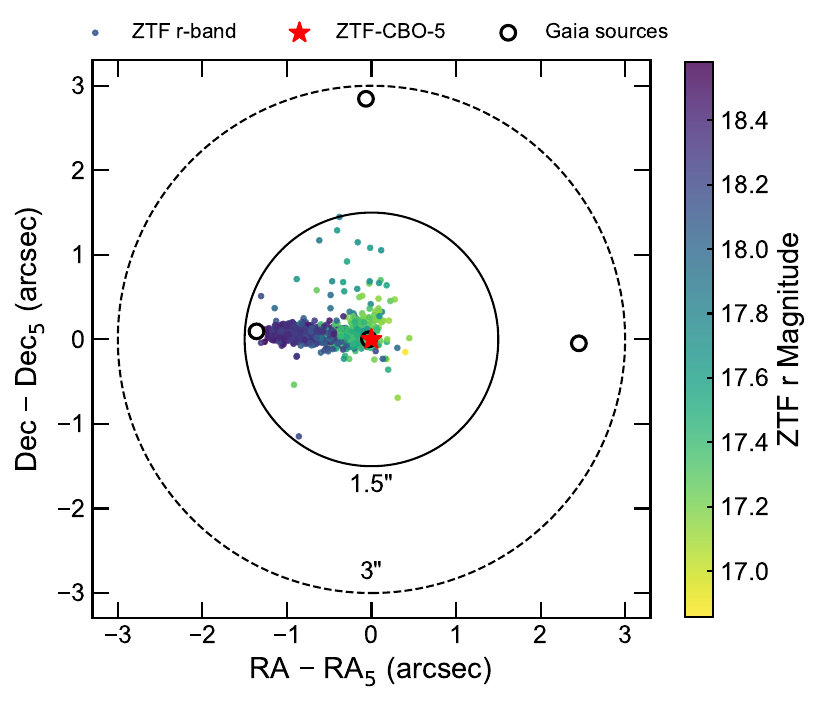}
\caption{Spatial distribution of ZTF $r$-band observations within 1.5$\arcsec$ of the centroid of ZTF-CBO-5, color-coded by magnitude. Nearby \gaia \ sources are marked by black circles, while the position of ZTF-CBO-5 is indicated by a red star. The spatial distribution of the observations correlates with $r$-band magnitude, where fainter detections tend to lie closer to the nearby contamination.}
\label{fig:DOC5_spatial}
\end{figure}

\begin{figure*}[t!]
\centering
\includegraphics[width=0.7\linewidth]{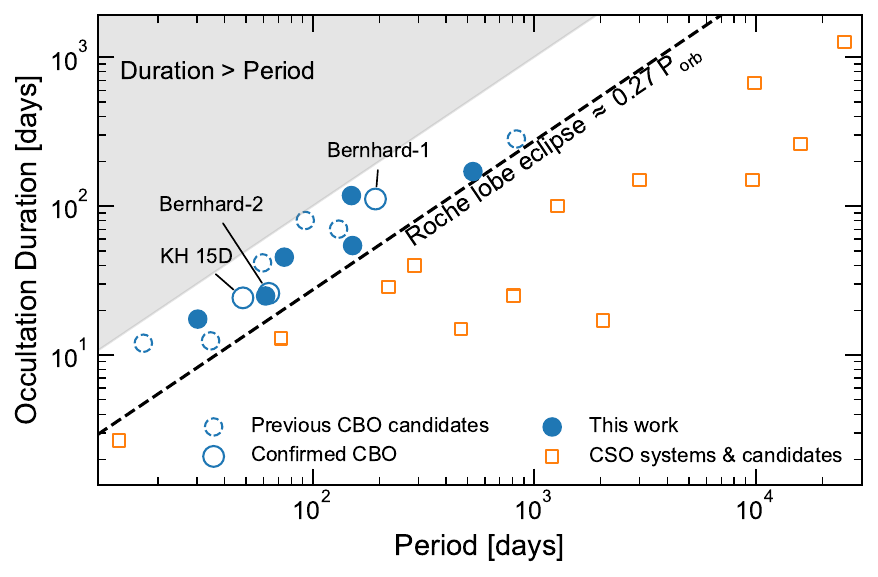}
\caption{Occultation duration versus orbital period for currently known CBO and CSO systems and candidates. Blue circles represent CBO systems: filled circles denote the six new candidates from this work, while solid and dashed circles correspond to confirmed and previously reported candidates, respectively. Orange squares indicate CSO systems, where occultation arises from material surrounding a companion. For KH~15D, we adopt a representative occultation duration of half the orbital period. The black dashed line indicates the theoretical maximum duration for an occultation induced by a Roche-lobe-filling disk around an equal-mass companion in a circular orbit (Equation~\ref{eq:roche_eq}). The shaded region indicates the unphysical regime where the occultation duration exceeds the orbital period. All CBO and CSO systems lie above and below the threshold, with a median orbital period of $\sim$74 days and $\sim 1277$ days, respectively.}
\label{fig:DO_sum}
\end{figure*}

\subsection{ZTF-CBO-5}

ZTF-CBO-5 exhibits the smallest amplitude among our candidates, with an out-of-occultation to occultation flux ratio of only $\sim$2 in ZTF $r$-band, as shown in Appendix \ref{app:A}. The system has an orbital period of $\sim$61 days and an occultation duration of $\sim$25 days (42\% of the cycle). The shallow depth is likely due to flux dilution from a nearby contaminating source located 1.5$\arcsec$ away, which is only 0.6 mag fainter than ZTF-CBO-5 in the G band.

This proximity introduces systematic effects in ZTF photometry, which has a resolution of $\sim$2$\arcsec$. During the bright phase, ZTF centroids lie closer to ZTF-CBO-5, while during the dark phase, they shift toward the contaminating source, as shown in Figure~\ref{fig:DOC5_spatial}. This could be understood as the blended point spread function follows the dominant flux contributor. The ZTF $r$-band magnitude during the bright phase ($\sim$17 mag) is roughly one magnitude brighter than concurrent PS observations, confirming that ZTF captures blended flux from both sources. 

The resulting blending introduces short timescale systematic flux variations with amplitudes comparable to the occultation depth, which significantly complicates the identification of the ingress. Consequently, the ingress crossing velocity $v_{\rm in}$ remains unconstrained in the light curve model posterior.

Despite this blending, independent PS observations confirm the occultation signature. With an angular resolution of $\sim$0.6$\arcsec$ \citep{Chambers2016_PS1, Flewelling2020_Pan-STARRS1}, PS is able to resolve the two sources. Panel c in Figure~\ref{fig:CBO5_P60} shows that PS detects periodic flux variations with roughly the same period and duration as the ZTF data, verifying that the occultation originates from ZTF-CBO-5.

\subsection{ZTF-CBO-6}

ZTF-CBO-6 exhibits the largest occultation to period ratio in our sample. The system has an orbital period of $\sim$150 days and an occultation duration of $\sim$118 days (77\% of the cycle). The ZTF and PS light curves are shown in Appendix~\ref{app:A}. A symmetric dip of approximately 7\% is also present near the center of the out-of-occultation baseline. Similar to ZTF-CBO-5, a nearby contaminating source 1.8$\arcsec$ away introduces systematic effects in ZTF observations, though less severe.

Both spatially filtered ZTF observations and independent PS data confirm the occultation signature. When restricting ZTF observations to within 0.5$\arcsec$ of the target coordinates, the periodic signature persists with consistent period and depth. The signature is also independently detected in the PS light curve.

\begin{deluxetable*}{ccccc}
\label{tab:DO_summary}
\tablecaption{Summary of CBO and CSO candidates}
\tablewidth{0pt}
\tabletypesize{\small}
\tablehead{
    \colhead{System Name} & 
    \colhead{Classification} & 
    \colhead{Period (days)} & 
    \colhead{Occultation Duration (days)} & 
    \colhead{Reference}
}
    
\startdata
CHS 7797 & CBO & 17.79(3) & $\sim$12 & (1) \\
ZTF-CBO-4 & CBO & 30.2566(4) & 17.43  & This work \\
ZTF J070412.91-112403.2 & CBO & $\sim$34.6 & $\sim$12 & (2) \\
KH 15D & CBO & 48.3777(2) & Variable, from zero to nearly full period & (3), (4) \\
VVV J180502.29-242501.4 & CBO & $\sim$59.35 & $\sim$41 & (5), (6) \\
ZTF-CBO-5 & CBO & 61.276(2) & 24.92  & This work \\
Bernhard-2 & CBO & 63.358(3) & Variable, around 20 & (7), (8) \\
ZTF-CBO-2 & CBO & 74.454(5) & 45.40  & This work \\
YLW 16A & CBO & 92.62(84) & $\sim$80 & (9) \\
WL 4 & CBO & 130.9(4) & $\sim$70 &  (10) \\
ZTF-CBO-6 & CBO & 149.62(1) & 117.59 & This work \\
ZTF-CBO-3 & CBO & 151.380(5) & 54.31 & This work \\
Bernhard-1 & CBO & 192.10(2) & Variable, abround 100 & (7), (11)\\
ZTF-CBO-1 & CBO & 529.32(3) & 170.42 & This work \\
VSSP J060413.69+300728.8 & CBO & $\sim$831 & $\sim$282 & VSX website \\
OGLE-LMC-ECL-17782 & CSO & $\sim$13.35 & $\sim$2.7 & (12) \\
MWC 882 & CSO & $\sim$72 & 12.96 & (13) \\
ELHC 10 & CSO & 219.9(24) & 28.59 & (14) \\
ZTF J1852+1249 & CSO & 289.57(9) & Variable, around 40 & (15) \\
OGLE-LMC-ECL-11893 & CSO & $\sim$468 & $\sim$15 & (16) \\
PDS 110 & CSO & 808(2) & $\sim$25 & (17) \\
OGLE-BLG182.1.162852 & CSO & $\sim$1277 & $\sim$100 & (18) \\
EE Cep & CSO & 2049.94 & $\sim$17 & (19) \\
eta Geminorum & CSO & $\sim$2993 & $\sim$150 & (20) \\
V773 Tau & CSO & $\sim$9672 & $\sim$150 & (21) \\
eps Aur & CSO & 9896.0(16) & $\sim$668 & (22) \\
ASASSN-24fw & CSO & $\sim$15987 & $\sim$261 & (23), (24) \\
TYC 2505-672-1 & CSO & $\sim$25221 & $\sim$1260 & (25) \\
\enddata
    
\tablecomments{
Systems are sorted by orbital period within the CBO and CSO categories. Uncertainties on occultation duration are omitted due to variability and inconsistent definitions across the literature. References: (1) \citet{Rodriguez-Ledesma2012_unusual}, (2) \citet{Bernhard2024_ZTFJ070412.91-112403.2}, (3) \citet{KH15D_discover}, (4) \citet{Winn2004_KH15D}, (5) \citet{Lucas2024_most}, (6) \citet{Rodriguez-Ledesma2012_CHS7797_spec}, (7) \citet{Zhu2022_Two}, (8) \citet{Hu2024_Bernhard2}, (9) \citet{Plavchan2013_identification}, (10) \citet{Plavchan2008_Peculiar}, (11) Hu et al. (in prep), (12) \citet{Graczyk2011_Optical}, (13) \citet{Zhou2018_Occultations}, (14) \citet{Garrido2016_eclipsing}, (15) \citet{Bernhard2024_ZTF}, (16) \citet{Dong2014_OGLE-LMC-ECL-11893}, (17) \citet{Osborn2017_Periodic}, (18) \citet{Rattenbury2015_OGLE-BLG182.1.162852}, (19) \citet{Galan2012_International}, (20) \citet{Torres2022_Geminorum}, (21) \citet{Kenworthy2022_Eclipsea}, (22) \citet{Carroll1991_Interpreting}, (23) \citet{Zakamska2025_ASASSN-24fw}, (24) \citet{Fores-Toribio2025_ASASSN-24fw}, (25) \citet{Rodriguez2016_Extreme}  
}
\end{deluxetable*}

\section{A Brief Overview of Periodic Long and Deep Eclipsing Objects}
\label{sec:do-summary}

The six new candidates reported here nearly double the known CBO sample, enabling meaningful population-level comparisons between CBO and CSO systems. We compiled periodic disk occultation candidates from Table 6 of \citet{Fores-Toribio2025_ASASSN-24fw}, the VSX catalog of E-DO systems\footnote{\hyperlink{https://vsx.aavso.org}{https://vsx.aavso.org}} \citep{Watson2006_International}, and a targeted literature search. After verifying reported properties against original publications and restricting to systems with at least two observed occultations, we identified 9 CBO and 13 CSO candidates with measured orbital periods (Table \ref{tab:DO_summary}).

The primary physical discriminant between CBO and CSO systems is the maximum occultation duration permitted by Roche lobe geometry. In the limiting case where circumstellar material fills the Roche lobe, the ratio of occultation duration to orbital period is
\begin{equation}
\frac{\tau_{\rm occult.}}{P_{\rm orb}} = \frac{\arcsin(R_{\rm L}/a_{2})}{\pi}
\label{eq:roche_eq}
\end{equation}
which yields $\approx 0.27$ for an equal-mass circular binary using Eggleton's approximation for Roche radius \citep{Eggleton1983_Roche}. As shown in Figure \ref{fig:DO_sum}, all current CBO candidates lie above this threshold, while all CSO candidates fall below it. All six new candidates exceed this limit, ruling out the model of the companion circumstellar materials eclipsing, and supporting their classification as CBO systems.

The median orbital period differs by an order of magnitude between CBO ($\sim$74 days) and CSO ($\sim$1277 days) systems, likely reflecting a detection bias. CBO systems have comparable durations for their dark and bright phases. When orbital periods approach the survey baseline on the order of $10^3$ days, only partial ingress or egress of the CBO systems may be captured, complicating period determination across surveys. CSO systems, with their brief occultations relative to orbital period, allow complete events to be observed and matched across surveys even at longer periods.

Other photometric properties, such as occultation depth, show no clear systematic trends between CBO and CSO populations. This is expected, as depth depends on both the central binary and the occulting material, which couples multiple physical effects that are difficult to disentangle from light curves alone.

Future time-domain surveys will likely expand this sample dramatically. For example, Vera C. Rubin Observatory (LSST) will monitor approximately ten times more stars than ZTF; given that ZTF has identified roughly ten CBO candidates, LSST could discover on the order of one hundred such systems. This wll provide unprecedented opportunities to probe misaligned disk physics and evolution. 

\section{Conclusion}
\label{sec:conclusion}

We report six new CBO candidates, namely ZTF-CBO-1 through ZTF-CBO-6, identified from a systematic search of ZTF photometry. These candidates display periodic brightness variations with substantial depth ($\gtrsim$1 mag) and long durations ($\gtrsim$30\% of their orbital periods). Their occultation-to-period ratios is large enough to rule out circumstellar disk interpretations and support their classification as CBO systems similar to KH~15D \citep{KH15D_discover,Winn2004_KH15D}, Bernhard-1, and Bernhard-2 \citep[][Hu et al. in prep]{Zhu2022_Two,Hu2024_Bernhard2}. The SEDs of ZTF-CBO-1 and ZTF-CBO-2 exhibit infrared excess, providing additional evidence for the presence of circumstellar dust.

A simple opaque-screen model reproduces the primary occultation features reasonably well, but failed to explain the photometric baseline modulation in ZTF-CBO-1 or the rapid ingress/egress variations in ZTF-CBO-3. Detailed modeling is limited by the lack of orbital constraints and sparse photometric sampling. Additional spectroscopic radial velocity follow-up and photometric monitoring are needed to fully characterize these systems.

These six candidates nearly double the known CBO sample, enabling population-level studies of misaligned circumbinary disks. Future surveys such as LSST, with $\sim$10 times more monitored stars, may discover on the order of one hundred such systems, providing unprecedented opportunities to probe disk-binary interactions and inner disk structure through occultation events.

\section*{acknowledgements}

We thank Zhen Guo for useful discussions. Work by Z.H. and W.Z. is supported by the National Natural Science Foundation of China (grant Nos. 12173021 and 12133005). Based on observations obtained with the Samuel Oschin Telescope 48-inch and the 60-inch Telescope at the Palomar Observatory as part of the Zwicky Transient Facility project. ZTF is supported by the National Science Foundation under Grants No. AST-1440341 and AST2034437 and a collaboration including current partners Caltech, IPAC, the Weizmann Institute for Science, the Oskar Klein Center at Stockholm University, the University of Maryland, Deutsches Elektronen-Synchrotron and Humboldt University, the TANGO Consortium of Taiwan, the University of Wisconsin at Milwaukee, Trinity College Dublin, Lawrence Livermore National Laboratories, IN2P3, University of Warwick, Ruhr University Bochum, Northwestern University and former partners the University of Washington, Los Alamos National Laboratories, and Lawrence Berkeley National Laboratories. Operations are conducted by COO, IPAC, and UW. This publication makes use of data products from the Two Micron All Sky Survey, which is a joint project of the University of Massachusetts and the Infrared Processing and Analysis Center/California Institute of Technology, funded by the National Aeronautics and Space Administration and the National Science Foundation. This publication makes use of data products from the Wide-field Infrared Survey Explorer, which is a joint project of the University of California, Los Angeles, and the Jet Propulsion Laboratory/California Institute of Technology, funded by the National Aeronautics and Space Administration. Some of the data presented herein were obtained at the W. M. Keck Observatory, which is operated as a scientific partnership among the California Institute of Technology, the University of California and the National Aeronautics and Space Administration. The Observatory was made possible by the generous financial support of the W. M. Keck Foundation. The Pan-STARRS1 Surveys (PS1) and the PS1 public science archive have been made possible through contributions by the Institute for Astronomy, the University of Hawaii, the Pan-STARRS Project Office, the Max-Planck Society and its participating institutes, the Max Planck Institute for Astronomy, Heidelberg and the Max Planck Institute for Extraterrestrial Physics, Garching, The Johns Hopkins University, Durham University, the University of Edinburgh, the Queen's University Belfast, the Harvard-Smithsonian Center for Astrophysics, the Las Cumbres Observatory Global Telescope Network Incorporated, the National Central University of Taiwan, the Space Telescope Science Institute, the National Aeronautics and Space Administration under Grant No. NNX08AR22G issued through the Planetary Science Division of the NASA Science Mission Directorate, the National Science Foundation Grant No. AST–1238877, the University of Maryland, Eotvos Lorand University (ELTE), the Los Alamos National Laboratory, and the Gordon and Betty Moore Foundation. This paper includes data collected with the TESS mission, obtained from the MAST data archive at the Space Telescope Science Institute (STScI). Funding for the TESS mission is provided by the NASA Explorer Program. STScI is operated by the Association of Universities for Research in Astronomy, Inc., under NASA contract NAS 5–26555. This work has made use of data from the European Space Agency (ESA) mission Gaia (https://www.cosmos.esa.int/gaia), processed by the Gaia Data Processing and Analysis Consortium (DPAC, https://www.cosmos.esa.int/web/gaia/dpac/consortium). Funding for the DPAC has been provided by national institutions, in particular the institutions participating in the Gaia Multilateral Agreement.

\vspace{5mm}
\facilities{ZTF, \gaia, 2MASS, Pan-STARRS, \tess, WISE}
\software{\texttt{astropy} \citep{astropy:2013,astropy:2018,astropy:2022}, 
          \texttt{scipy} \citep{2020SciPy-NMeth}, 
          \texttt{emcee} \citep{Foreman-Mackey2013_emcee}, 
          \texttt{isochrones} \citep{Morton2015_isochrones}, 
          \texttt{tglc} \citep{Han2023_TESS-Gaia}
          }

\newpage
\appendix

\section{Light curves of ZTF-CBO-3 to 6}
\label{app:A}

\begin{figure*}
\centering
\includegraphics[width=0.9\linewidth]{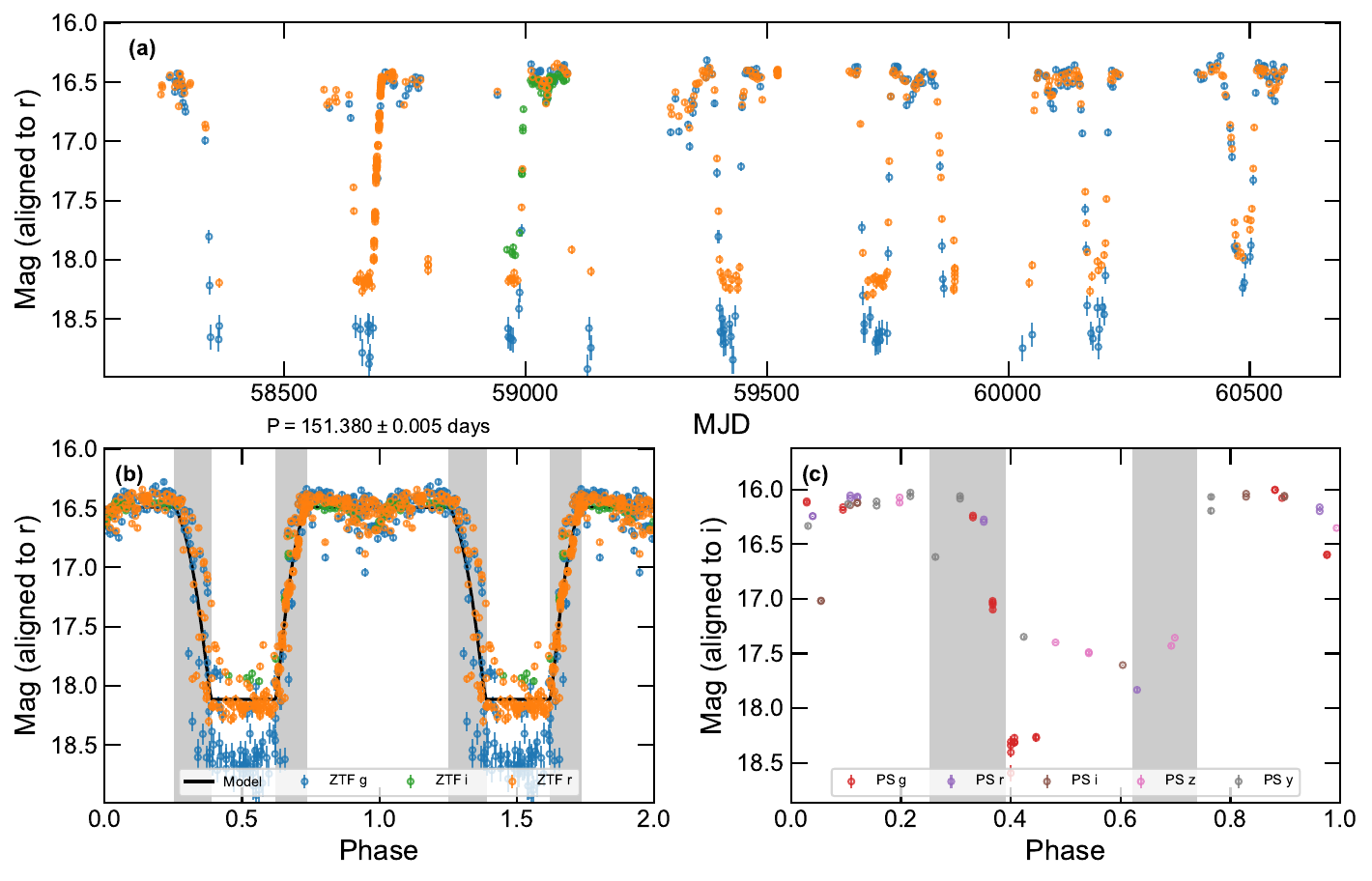}
\caption{Same as Figure \ref{fig:CBO1_P530}, but for ZTF-CBO-3. }
\label{fig:CBO3_P150}
\end{figure*}

\begin{figure*}
\centering
\includegraphics[width=0.9\linewidth]{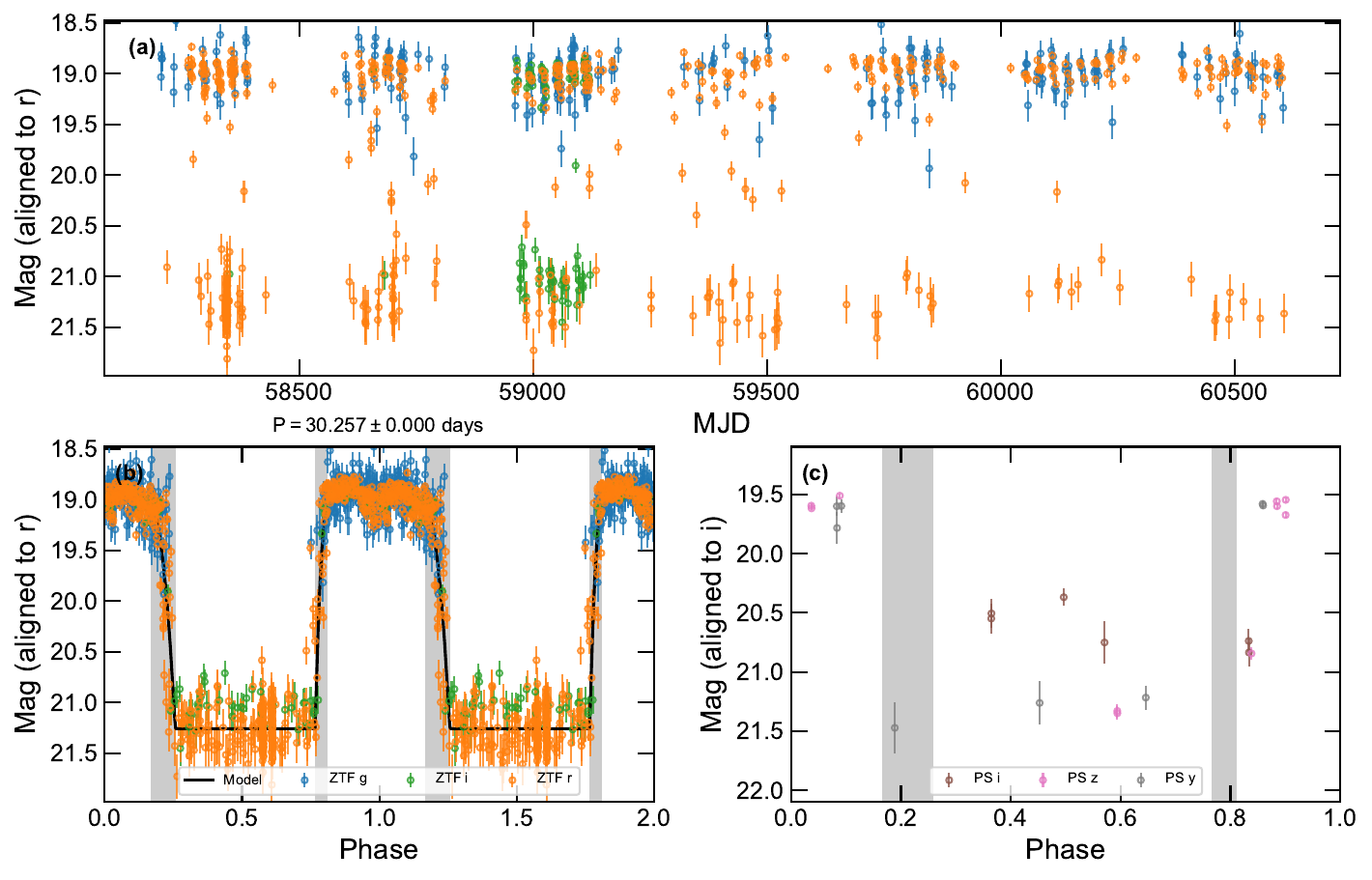}
\caption{Same as Figure \ref{fig:CBO1_P530}, but for ZTF-CBO-4.}
\label{fig:CBO4_yso_P30}
\end{figure*}

\begin{figure*}
\centering
\includegraphics[width=0.9\linewidth]{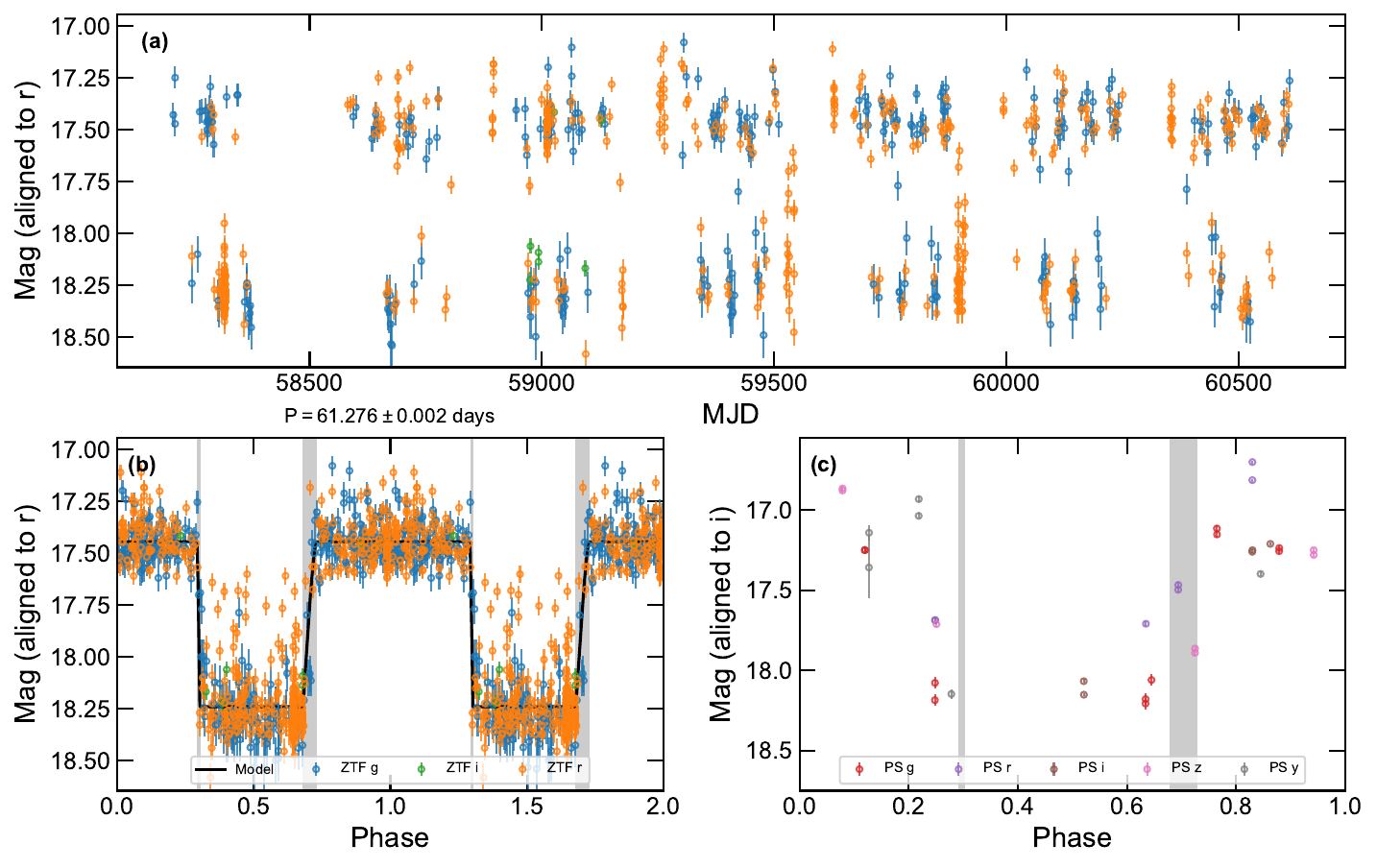}
\caption{Same as Figure \ref{fig:CBO1_P530}, but for ZTF-CBO-5. From panel b, we can see that there are many outliers above the average brightness of the bright and the dark phase. We cannot rule out the possibility that these observations are biased by systematics.}
\label{fig:CBO5_P60}
\end{figure*}

\begin{figure*}
\centering
\includegraphics[width=0.9\linewidth]{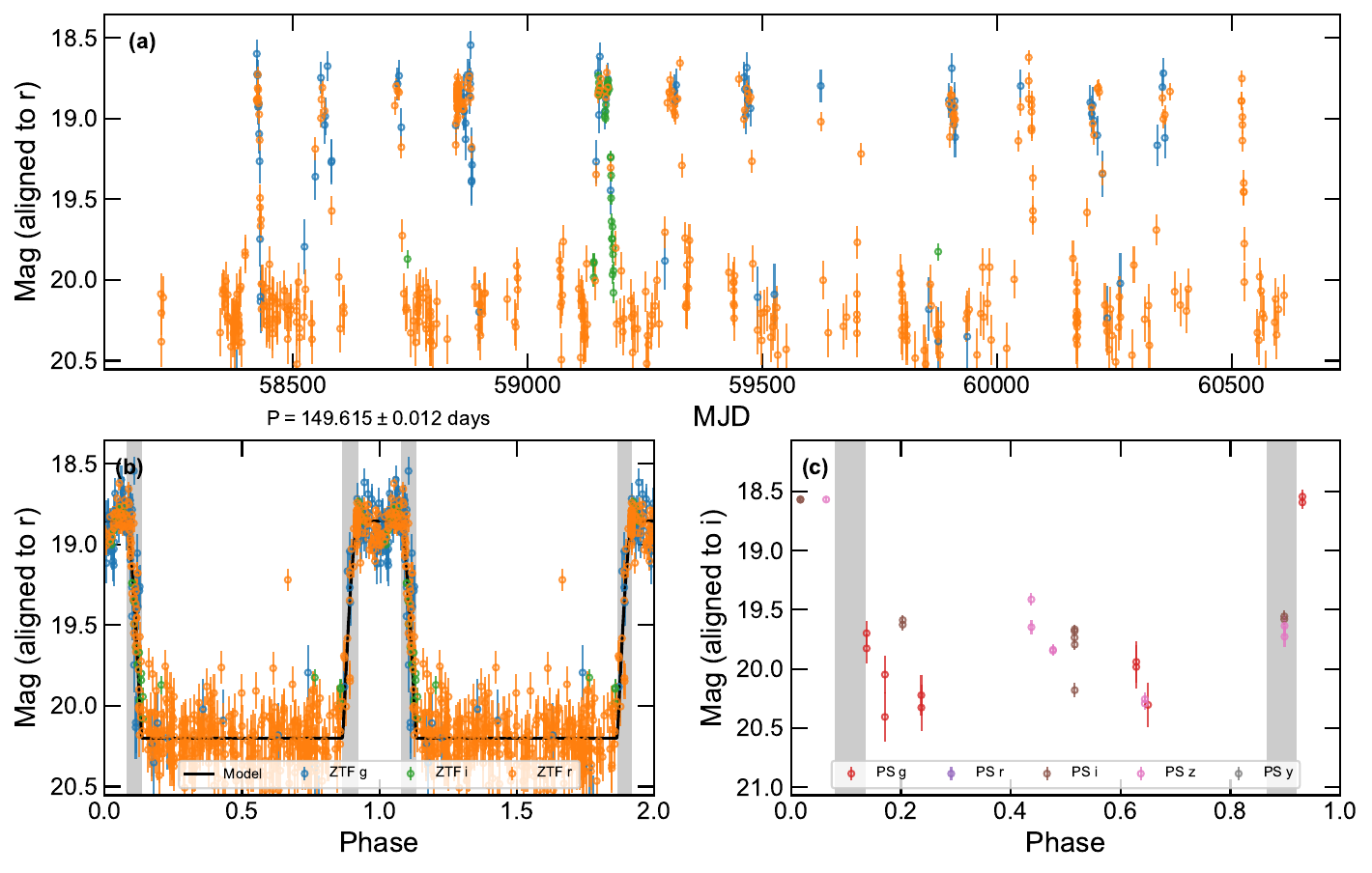}
\caption{Same as Figure \ref{fig:CBO1_P530}, but for ZTF-CBO-6.}
\label{fig:CBO6_P150}
\end{figure*}

Here we show the original ZTF light curve, phase folded ZTF light curve, and phase folded PS light curve for ZTF-CBO-3, 4, 5, and 6, in Figure \ref{fig:CBO3_P150}, \ref{fig:CBO4_yso_P30}, \ref{fig:CBO5_P60}, and \ref{fig:CBO6_P150}, respectively.

\bibliography{sample631}{}
\bibliographystyle{aasjournal}

\end{document}